\documentclass[conference]{IEEEtran}

\usepackage[linesnumbered,lined, algoruled]{algorithm2e}
\usepackage{amssymb,bm}
\usepackage{amsmath}
\usepackage[acronym,shortcuts]{glossaries}
\usepackage{algorithmic,float}
\usepackage{balance}
\usepackage{color}
\usepackage{dsfont}
\usepackage{soul}
\usepackage{graphicx}
\usepackage{comment}
\makeglossaries
\newacronym{5g}{5G}{fifth generation}
\newacronym{aod}{AOD}{angle of departure}
\newacronym{awgn}{AWGN}{additive white Gaussian noise}
\newacronym{bler}{BLER}{block error rate}
\newacronym{bs}{BS}{base station}
\newacronym{ecf}{ECF}{empirical characteristic function}
\newacronym{cdf}{CDF}{cumulative density function}
\newacronym{cqi}{CQI}{channel quality indicator}
\newacronym{ce}{CE}{cross entropy}
\newacronym{cp}{CP}{cyclic prefix}
\newacronym{harq}{HARQ}{hybrid automatic repeat request}
\newacronym{ipv}{IPV}{interference power value}
\newacronym{kl}{KL}{Kullback-Leibler}
\newacronym{ks}{KS}{Kolmogorov-Smirnov}
\newacronym{la}{LA}{link adaptation}
\newacronym{los}{LOS}{line of sight}
\newacronym{mac}{MAC}{medium access control}
\newacronym{mbb}{MBB}{mobile broadband}
\newacronym{ml}{ML}{machine learning}
\newacronym{mcs}{MCS}{modulation and coding scheme}
\newacronym{mimo}{MIMO}{multiple-input multiple-output}
\newacronym{mp}{EB}{expectation based}
\newacronym{mq}{MQ}{maximum quantile}
\newacronym{mrt}{MRT}{maximum ratio transmission}
\newacronym{mse}{MSE}{mean squared error}
\newacronym{ofdm}{OFDM}{orthogonal frequency division multiplexing}
\newacronym{olla}{OLLA}{outer loop link adaptation}
\newacronym{pdf}{PDF}{probability density function}
\newacronym{pfs}{PFS}{proportional fair scheduling}
\newacronym{ppp}{PPP}{Poisson point process}
\newacronym{rr}{RR}{round robin}
\newacronym{rv}{r.v.}{random variable}
\newacronym{se}{SE}{spectral efficiency}
\newacronym{sinr}{SINR}{signal to interference plus noise ratio}
\newacronym{sir}{SIR}{signal to interference ratio}
\newacronym{snr}{SNR}{signal to noise ratio}
\newacronym{svm}{SVM}{support vector machine}
\newacronym{tdd}{TDD}{time division duplex}
\newacronym{tti}{TTI}{transmission time interval}
\newacronym{ue}{UE}{user equipment}
\newacronym{ula}{ULA}{uniform linear array}
\newacronym{urllc}{URLLC}{ultra-reliable low-latency communications}

\begin{document}
\title{Interference Distribution Prediction for Link Adaptation in Ultra-Reliable Low-Latency Communications}

\author{Alessandro Brighente$^{*\mathsection}$, Jafar Mohammadi$^\mathsection$, and Paolo Baracca$^\mathsection$\\
$^*$Department of Information Engineering, University of Padova, Italy\\
$^\mathsection$Nokia Bell Labs, Stuttgart, Germany}

\maketitle

\sloppy
\begin{abstract}
The strict latency and reliability requirements of \ac{urllc} use cases are among the main drivers in \ac{5g} network design.
\Ac{la} is considered to be one of the bottlenecks to realize URLLC.
In this paper, we focus on predicting the signal to interference plus noise ratio at the user to enhance the \ac{la}.
Motivated by the fact that most of the \ac{urllc} use cases with most extreme latency and reliability requirements are characterized by semi-deterministic traffic, we propose to exploit the time correlation of the interference to compute useful statistics needed to predict the interference power in the next transmission.
This prediction is exploited in the \ac{la} context to maximize the spectral efficiency while guaranteeing reliability at an arbitrary level.
Numerical results are compared with state of the art interference prediction techniques for \ac{la}. We show that exploiting time correlation of the interference is an important enabler of \ac{urllc}.
\end{abstract}


\glsresetall

\section{Introduction}\label{sec:intro}
During the past decades, communication networks have been engineered to be human-centric, targeting network capacity and assuming a small number of users. With the \ac{5g} of mobile networks, an increasing amount of traffic data types is supposed to be supported by the same networks. Among these heterogeneous types of traffic, \ac{urllc} poses one of the major challenges in the design of both physical and medium access control layer. This type of traffic is characterized by the fact that transmissions should be performed at almost zero delay (latency is targeted to be $1$~ms or less), while ensuring very high reliability (failure probability targeted to $10^{-5}$ or less) for packet of few tens of bytes \cite{3gpp.22.104}. With \ac{mbb}, reliability is achieved via re-transmission of packets with techniques such as hybrid automatic repeat request. However, reliability can not be guaranteed for \ac{urllc} with several re-transmissions, as that implies an overhead in time that does not allow meeting the strict latency constraint. These targets require, therefore, a change of view in both the way reliability is obtained and the way that scheduling is performed in typical wireless scenarios. Among the solutions that have been proposed for low-latency reduction, we find the use of short packets \cite{durisi2016} and the use of shorter \acp{tti} \cite{li20175g} very promising. We refer the reader to \cite{bennis2018} for a complete survey on the characteristic of \ac{urllc} related techniques.

In cellular networks, \ac{la} is responsible for the choice of the \ac{mcs} based on the observed channel condition. In the ideal setting, the \ac{mcs} that maximizes the \ac{se} while guaranteeing a target \ac{bler} will be chosen. 
\ac{la} has been shown in \cite{shariatmadari2016} to be an effective tool to insure a given reliability.
It is important to note that \ac{urllc} traffic is characterized by short packets, which occupy a subset of the radio resources within a \ac{tti}. The result is that the interference pattern changes rapidly: it is then difficult to select a proper \ac{mcs} just based on the observed \ac{cqi}, which becomes outdated very quickly. Traditionally \ac{la} has been addressed via \ac{olla} \cite{sampath1997}. Although working well when at least one or few re-transmissions are allowed \cite{pocovi2018}, \ac{olla} leads to a very conservative behaviour with related waste of resources when \ac{urllc} latency constraints become extreme and do not allow re-transmissions.

The rapid change in the interference pattern with \ac{urllc} traffic has indeed been considered in \cite{pocovi2018}, where authors propose to low pass filter the currently estimated \ac{sinr} with that estimated at the previous time instant: the authors show that extending \ac{olla} with this improvement allows serving \ac{urllc} traffic when one re-transmission is allowed. On the other hand, semi-deterministic periodic traffic characterizes many factory automation \ac{urllc} use cases, for instance the cyber-physical control applications \cite[Sect. 5.2]{3gpp.22.104}, which are characterized by extreme latency and reliability requirements. As a consequence, although the interference pattern changes very quickly in these scenarios, the traffic periodicity is reflected also in the interference suffered by a certain \ac{ue}. In our work, we propose to exploit the time correlation in the interference pattern by computing the conditional distribution of successive interference values in order to predict the interference at the successive time instant. The main focus is on the estimation method of such distributions, and on the predictive methods which allow to control the reliability of the system. 

This concept has also been investigated in \cite{oruthota2016}, where \ac{la} is performed based on the statistics of the \ac{sinr} assuming Rayleigh small scale fading and Jakes' model for the Doppler spectrum. However, that proposed solution is model dependent and does not exploit the time correlation of the interference pattern.

In \cite{levanen2015} interference is measured at a central \ac{bs} and interferers are grouped into concentric layers. \acp{ue} are assumed to be static, time correlation of fast fading is modelled by mean of a Bessel function of the first kind and traffic is generated according to an interrupted Poisson process. 

In \cite{taranetz2015} the circular interference model is proposed, where interferers are clustered in concentric circles, one for each \ac{ue} which causes strong interference. Other \acp{ue} are then modelled by a power profile for different locations in each circle, and interference power is modelled as a summation of Gamma \acp{rv}. The accuracy of the circular interference model is evaluated by mean of the Kolmogorov-Smirnov distance between the \ac{cdf} of the circular model and the true \ac{cdf}.

Rayleigh fading channels are considered in \cite{feng2019}, and shadowing is considered as noise source. \acp{bs} locations are assumed to follow a Poisson point process and user coordination is obtained by assigning each user to the nearest \ac{bs}. Defining as success probability the probability of the signal to interference ratio being above a certain threshold, \cite{feng2019} provides the distribution of the success probability conditioned on \ac{ue} being a cell center or cell boundary type.

Finally, in \cite{zhuang2011} a geometrical-based probability model is proposed to model the distribution of the distance between \acp{ue} and \ac{bs} in a cellular network. Then, considering a channel affected only by path-loss, a closed form expression for the distribution of the interference caused by neighbouring cells is derived.

Against this background, we propose a prediction method which does not rely on any particular channel model. Therefore it can be applied without loss of generality to any channel condition and scenarios. We show how interference correlation in time is a useful characteristic which can be exploited to predict transmission \ac{sinr} for \ac{la} purposes. We derive two methods for interference prediction, namely the \ac{mp} and the \ac{mq}, and we show that with \ac{mq} we are able to control the reliability level of the system while maximizing the experienced \ac{se}. Results are compared with the state of the art interference prediction for \ac{la}, and results show that exploiting the statistical characterization of the interference is an important enabler for \ac{urllc}.  

\section{System Model}
\begin{figure}
  \centering
    \includegraphics[width=0.5\textwidth]{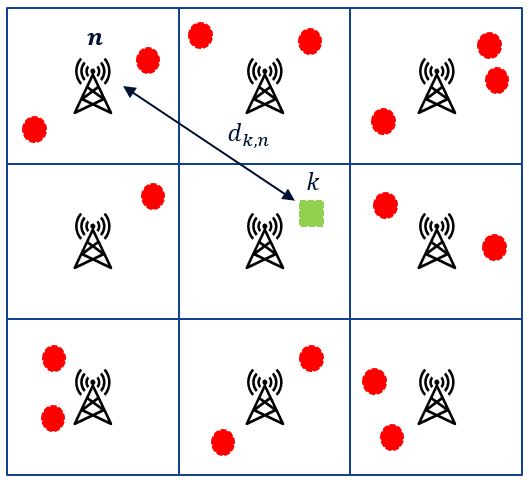}
    \caption{Example of the considered two-dimensional scenario with $N=9$ \acp{bs}.}
    \label{fig:2d}
\end{figure}

We consider a cellular network with $N$ cells, wherein a single \ac{bs} equipped with $N_a$ antennas is located at the center of each cell, as shown in Fig. \ref{fig:2d}. Each cell is populated by a random number of single-antenna \acp{ue} uniformly distributed between $K_{\rm min}$ and $K_{\rm max}$ and uniformly located in space. We denote with $K_{\rm tot}$ the total number of \acp{ue} in the networks, and with $\mathcal{X}_n$ the set of indexes of the \acp{ue} located in cell $n$. Furthermore we define as $(x^n,y^n)$, $n=1,\cdots, N$ and $(v^k,z^k)$, $k=1,\cdots, K_{\rm tot}$ respectively the coordinates of the locations of \acp{bs} and \acp{ue}. Assuming that both \acp{ue} and \acp{bs} are at the ground level, the distance between \ac{bs} $n$ and \ac{ue} $k$ is given by
\begin{equation}
d_{k,n} = \sqrt{(x^n-v^k)^2+(y^n-z^k)^2},
\end{equation}
with corresponding \ac{aod}
\begin{equation}
\theta_k =
\sin^{-1}\left( \frac{z^k-y^n}{d_{k,n}}\right) .
\end{equation}

\subsection{Channel Model}

Assuming that antennas at the \acp{bs} are organized in a \ac{ula}, the gain of the transmission toward \ac{aod} $\theta_k$ is given by vector $\bm{a}^{n}$, whose $m$~th component is \cite{balanis2016}
\begin{equation}
 [\bm{a}^{n}(\theta_k)]_m = e^{j \delta_k}\exp\left(\frac{j 2 \pi \lambda (m-1) \cos(\theta_k)}{d}\right),
\end{equation}
where $[\bm{a}^{n}]_m$ denotes the $m-$~th element of vector $\bm{a}^{n}$, $m=1, \cdots, N_a$, $\lambda$ denotes the carrier wavelength, $d$ is the inter antenna spacing, $\delta_k$ is a user specific phase shift, and $j = \sqrt{-1}$.

The \ac{los} component of the channel between the $n-$~th \ac{bs} and \ac{ue} $k$ is given by
\begin{equation}
	\bm{h}_{\rm LOS}(k,n) = \sqrt{\frac{K_0}{d_{k,n}^{\nu}}}\bm{a}^{n}(\theta_k),
\end{equation}
where $\nu$ is the path loss exponent and $K_0$ is a constant factor accounting for the cell edge \ac{snr}. 

We assume the Rice model for the small scale fading, and the channel can be modeled as the sum of a \ac{los} path and a random multi-path component, i.e., \cite[Ch. 2.4.2]{tse2005}
\begin{equation}
 \bm{h}(k,n) = \sqrt{\frac{\Psi}{\Psi+1}} \bm{h}_{\rm LOS}(k,n) + \sqrt{\frac{1}{\Psi+1}} \bm{h}_{\rm NLOS}(k,n),
\end{equation}
where $\Psi$ denotes the Rice factor and $\bm{h}_{\rm NLOS}(k,n)\sim\mathcal{CN}\left(\bm{0},K_0/d_{k,n}^{\nu}\bm{I}\right)$, with $\bm{0}$ and $\bm{I}$ being a vector with $N_a$ zero entries and the $N_a \times N_a$ identity matrix, respectively.

We assume that each \ac{bs} serves a single \ac{ue} per time instant by using a \ac{mrt} beamformer, and we denote as $\bm{g}(k,n)$ the \ac{mrt} beamformer from \ac{bs} $n$ toward \ac{ue} $k$, given by
\begin{equation}
\bm{g}(k,n) = \frac{\bm{h}^H(k,n)}{||\bm{h}(k,n)||},
\end{equation}
where $[\cdot]^H$ denotes the Hermitian of a vector.

\ac{bs} $n$ transmits a symbol $q_k$ such that $\mathbb{E}[|q_k^2|] =  1$ to \ac{ue} $k$, and the received signal is expressed as
\begin{equation}
 y_{k,n} = \bm{h}(k,n)\bm{g}(k,n)q_k\sqrt{P} + w,
\end{equation}
where $P$ is the transmitted power and $w$ is the \ac{awgn} component distributed as a Gaussian \ac{rv} with zero mean and variance $\sigma^2$.

The signal received by user $k$ suffers the interference caused by all \acp{bs} $\ell \neq n$  transmitting toward their scheduled \acp{ue}. The received \ac{sinr} by \ac{ue} $k$ is given by 
\begin{equation}\label{eq:sinr}
\rho_{k,n} = \frac{|\bm{h}(k,n)\bm{g}(k,n)|^2P}{\sum_{\ell = 1, \ell \neq n}^{N}|\bm{h}(k,\ell)\bm{g}(x_\ell,\ell)|^2P+ \sigma^2},
\end{equation}
where $x_\ell$ denotes the index of the \ac{ue} served by the $\ell$-th \ac{bs}. 


In this work, we consider two different \ac{ue} scheduling policies, i.e., either \ac{rr} or \ac{pfs}. When considering the semi-deterministic periodic traffic, we can assume a \ac{rr} scheduling of the \acp{ue} at each \ac{bs}, with a) the order properly selected in order to allow each \ac{ue} to meet its latency requirement, and b) a  fully loaded \ac{bs} in terms of served \acp{ue}, i.e., in each \ac{tti} there is always a certain UE that needs to be scheduled by each BS. Therefore, with this \ac{rr} scheduling, \acp{ue} in $\mathcal{X}_n$ are served in a deterministic fashion and, without loss of generality, we assume that they are sequentially served according to their index in $\mathcal{X}_n$. Note that, with these modeling assumptions, packets that are successfully received at the \acp{ue} always meet the latency constraints: on the other hand, when a packet transmission fails because of a wrong \ac{sinr} prediction, the packet is just dropped, and that negatively affects the system reliability.

On the other hand, \ac{pfs} \cite{viswanath2002} does not necessarily serve \acp{ue} in a deterministic fashion, and thus does not represent the most suitable solution for \ac{urllc} with extreme latency constraint. However, by guaranteeing the same long term average throughput, it has been widely used for \ac{mbb} and its variations are envisioned to be used also for \ac{urllc} with not too extreme latency requirements \cite{sisinni2016}. Note that, to ensure \ac{ue} fairness, \ac{pfs} serves \acp{ue} in a certain \ac{tti} that have not been served in a while and therefore introduces some time periodicity in the interference power experienced by the \acp{ue}, in particular when the number of \acp{ue} per cell is not too high. 

\subsection{Link Adaptation}\label{sec:laIntro}
\ac{la} is a technique used to control \acp{ue} quality of service in which, according to the channel condition, the \ac{bs} chooses the best \ac{mcs} in order to match a target \ac{bler}. Denoting as $\Omega$ the set of available \ac{mcs} indexes, the \ac{la} problem for \ac{ue} $k$ served by \ac{bs} $n$ can be formulated as
\begin{subequations}\label{eq:la}
\begin{equation}
\omega_k^* = \underset{\omega\in\Omega}{\arg \max} \, R_k(\omega|\hat{\rho_{k,n}})
\end{equation}
\begin{equation}
\text{s.t.} \quad \Gamma(\omega|\hat{\rho_{k,n}}) \leq \Gamma_0,
\end{equation}
\end{subequations}
where $\Gamma(\omega|\hat{\rho_{k,n}})$ and $R_k(\omega|\hat{\rho_{k,n}})$ are respectively the \ac{bler} and supported data rate given the estimated \ac{sinr} $\hat{\rho_{k,n}}$ for user $k$ obtained choosing \ac{mcs} $\omega$, and $\Gamma_0$ is the target \ac{bler}.

The basic \ac{la} algorithm is the \ac{olla} where the estimated \ac{sinr} value is modified based on the reception of the previously sent packets \cite{sampath1997}. 
Therefore, in order to guarantee high reliability with very low latency with \ac{olla}, for a long period of time the estimated \ac{sinr} used for \ac{la} will be much lower than the actual one, i.e., $\hat{\rho_{k,n}} << \rho_{k,n}$. The result is that reliability is obtained at the cost of wasting resources in terms of throughput. The \ac{olla} algorithm is effective only when the target of the communication is not ultra reliability (for which the target \ac{bler} is set to $\Gamma_0 = 10^{-5}$, or even lower), or when one or few re-transmissions are allowed. 

We hence note that, in order to guarantee a certain \ac{bler} while at the same time exploiting all the resources in terms of throughput, we need a good estimate of the \ac{sinr}. 

In this paper, we propose a novel approach to deal with the choice of the \ac{mcs} based on the statistics of the \ac{sinr} and in particular on the statistics of the \acp{ipv}, exploiting the time correlation between successive \acp{ipv}. We target the maximization of the \ac{se} while ensuring a certain target \ac{bler}. A mathematical formulation of the problem will be given in the following sections.

\section{Interference Power Estimation}\label{sec:intEst}
In order to choose the most suitable \ac{mcs}, we need to be able to predict the behavior of the channel. In particular, considering that the received power from the serving cell varies in a much lower time scale when compared to the interference \cite{holma2012}, we are interested in the estimation of the interference power. Considering a transmission from \ac{bs} $n$ to \ac{ue} $k$, the \ac{ipv} random variable $I_k$ is given by the first term at the denominator in (\ref{eq:sinr}), i.e.,
\begin{equation}
I_k = \sum_{\ell = 1, \ell \neq n}^{N}|\bm{h}(k,\ell)\bm{g}(x_\ell,\ell)|^2P. 
\end{equation}

Since the IPV depends on the channel, which is a time dependent random process, for each given time instance $t$ the IPV can be represented as a random variable, i.e., $I_k(t)$. 
For a certain \ac{ue} $k$, we consider $L$ \acp{ipv} in successive time instants, and we define the set $\mathcal{P}_k = \{ I_k(1), \cdots, I_k(L)\}$ as the set of \acp{ipv} for \ac{ue} $k$ during a time period $L$. 
The set $\mathcal{P}_k$ can be seen as an ergodic time series from which we can compute useful statistics for \ac{la} purposes.
In the following, for the sake of clarity, we skip the subscript $k$ from the IPV random variable $I(t)$.

In particular, we are interested in the distribution of the \ac{ipv} at time instant $t+1$ given $D$ previous \acp{ipv} observations, i.e., in the conditional probability $\mathbb{P}\left(I(t+1)|\bm{I}_D(t)\right)$, where $\bm{I}_D(t)=[I(t),\cdots,I(t-D+1)]$. This probability can be computed as
\begin{equation}\label{eq:condPDF}
\mathbb{P}\left(I(t+1)|\bm{I}_D(t)\right) = \frac{\mathbb{P}\left(I(t+1),\bm{I}_D(t)\right)}{\mathbb{P}\left(\bm{I}_D(t)\right)}. 
\end{equation}

The marginal and joint distributions in (\ref{eq:condPDF}) can respectively be computed from set $\mathcal{P}_k$ by shifting a length $D$ and $D+1$ window over the length $L$ series, where in the latter the sample at $D+1$ is the value to be predicted, i.e., that at $t+1$.  

The simplest approach to estimate the \acp{pdf} for each \ac{ue} $k$ is to compute the empirical \ac{cdf} $F_k(\bm{i})$ of $\mathcal{P}_k$ as
\begin{equation}\label{eq:empCDF}
F_k(\bm{i}) = \frac{\sum_{t=D}^L \mathds{1}(\bm{I}_D(t) < \bm{i})}{L-D+1},
\end{equation}
where $\mathds{1}(\bm{I}_D(t) < \bm{i})$ is the counter of the number of vectors in $\mathcal{P}_k$ smaller than $\bm{i}$, where the inequality is component-wise. The empirical \ac{pdf} is then derived from the CDF. 
However this method has the drawback  that, if $L$ is not large enough, the entries with lower probability value will not appear in $\mathcal{P}_k$, and therefore we 
assume value $0$ for such entries. This problem becomes more prominent in the case of \ac{urllc}, since the extreme conditions that provide reliability with error probability of the very small magnitudes, e.g. $10^{-5}$, will be of focus. Using the histogram for such an application only makes sense if $L$ is large enough such that there are entries in the low probability part.  
In the next section we review the kernel based distribution estimation approach, which fills the gaps of the previously described distribution estimators.

\section{Kernel Based PDF Estimation}
We adopt the approach proposed in \cite{botev2010} to compute the joint \ac{pdf} of the distribution of sets of $D$ and $D+1$ \acp{ipv} measured over consecutive time instants. These samples are filtered with the kernel function $\mathcal{K}$, from which we obtain the \ac{pdf} \cite{botev2010}
\begin{equation}
f_k(\bm{i}) = \frac{1}{L}\sum_{t=D}^L \mathcal{K}\left(\bm{i},\bm{I}_D(t),\gamma\right),
\end{equation}

where $\bm{I}_D(t)$ is the $t$-th $D \times 1$ size vector of successive \acp{ipv}, $\bm{i}$ is the $D \times 1$ vector for which we want to compute the \ac{pdf}, and $\gamma$ is an optimization parameter defined as the \textit{bandwidth}. In order to compute the optimal bandwidth we exploit the key observation of \cite{botev2010} that the Gaussian kernel
\begin{equation}
\mathcal{K}\left(\bm{i},\bm{I}_D(t),\gamma\right) = \frac{1}{\sqrt{2 \pi \gamma}} exp \left(-\frac{(\bm{i}-\bm{I}_D(t))^H(\bm{i}-\bm{I}_D(t))}{2\gamma}\right)
\end{equation}
is the solution of the diffusion partial differential equation 
from which the optimal bandwidth is computed. Please refer to \cite{botev2010} for further details. 

Once both $f_k\left(I(t+1),\bm{I}_D(t)\right)$ and $f_k\left(\bm{I}_D(t)\right)$ have been computed, we use (\ref{eq:condPDF}) to obtain the conditional \ac{pdf}
\begin{equation}
f_k\left(I(t+1)|\bm{I}_D(t)\right) = \frac{f_k\left(I(t+1),\bm{I}_D(t)\right)}{f_k\left(\bm{I}_D(t)\right)}. 
\end{equation}
In the next section, we propose different approaches which can be used for interference prediction.

\section{Interference prediction for link adaptation}

Given the \ac{la} optimization problem (\ref{eq:la}), as in Section \ref{sec:laIntro}, the choice of the \ac{mcs} is based on the estimated \ac{sinr}, and in particular on the predicted \ac{ipv}. Based on this consideration, we rewrite the \ac{la} problem as

\begin{align} \nonumber
\min_{\hat{I}(t+1)} & \quad \hat{I}(t+1) \\
\text{s.t.}& \quad \mathbb{P}\left(\hat{I}(t+1)  < I(t+1)|\bm{I}_D(t) \right) \leq \Gamma_0,\label{eq:condCDF}
\end{align}
where $\hat{I}(t+1)$ is the predicted \ac{ipv}. The motivation of (\ref{eq:condCDF}) is given in the following. Since we require to estimate the \ac{ipv} for each time instant, in order to maximize the rate we choose the highest possible \ac{mcs}. Thus we are interested in minimizing the value of the estimated \ac{ipv}, while fulfilling the reliability requirement. According to the estimated \ac{sinr} value, the \ac{bs} chooses an \ac{mcs}, which is assumed to cause an error if the estimated \ac{sinr} is higher than the actual one. Therefore, as per our assumed model, the only source of uncertainty is the estimated \ac{ipv}, i.e., an error occurs if the estimated \ac{ipv} is below the actual one both computed at the successive time instant.

\textbf{Remark:} As opposed to the prediction methods in the literature in Section \ref{sec:intro}, the proposed method does not rely on the channel model and can be used in all those scenarios for which traffic and therefore interference is assumed to have some time correlation.


\subsection{Expectation Based Prediction}
In the \ac{mp} prediction framework the choice of the interference power value to be used for \ac{la} is made by choosing the expected value given the previous $D$ observations of the interference power, i.e.,
\begin{equation}
I^{\rm (EM)}(t+1) = \mathrm{E}\left[I(t+1)|\bm{I}_D(t)\right],
\end{equation}
where the expectation is computed empirically.
Although simple, this approach has the drawback that it does not allow to control the \ac{bler} and in particular to upper bound it. Therefore, by exploiting the statistical knowledge of the conditional distribution of the successive \acp{ipv}, we propose the more conservative \ac{mq} framework.

\subsection{Maximum Quantile Prediction}
A more conservative way of predicting the successive \ac{ipv} and at the same time control the probability of error is obtained by considering a certain quantile of the \ac{cdf} obtained by the conditional probability of the successive \acp{ipv}. In particular, we consider the fact that a transmission error occurs when the estimated \ac{sinr} is above the actual one, and therefore, as we focus on \ac{ipv} estimation, an error occurs when the estimated \ac{ipv} is below the actual one.

On one hand, we aim at both minimizing the selected \ac{ipv} in order to maximize the throughput and fulfilling the reliability constraint without choosing an unnecessarily low \ac{ipv}. On the other hand, the CDF function in the (\ref{eq:condCDF}) is a monotonically decreasing function of $\hat{I}(t+1)$ by definition, thus the optimal solution shall fulfill (\ref{eq:condCDF}) with equality:
  
\begin{subequations}
\begin{equation}
I^{\rm (MQ)}(t+1) = \hat{I}(t+1)
\end{equation}
\begin{equation}
\text{s.t.} \quad \mathbb{P}\left( \hat{I}(t+1) < I(t+1)|\bm{I}_D(t)\right) = \Gamma_0.
\end{equation}
\end{subequations}

\section{Numerical Results}
We here compare the results achieved with the proposed \ac{mp} and \ac{mq} heuristics against some state of the art techniques. We consider a scenario with $N=9$ cells, where the distance between neighboring \acp{bs} is $200$~m, and each \ac{bs} is equipped with $N_a=16$ antennas linearly spaced by $d=\lambda/2$. The \acp{ipv} are measured in the central cell, and the number of \acp{ue} in the surrounding cells is distributed between $K_{\rm min} =2$ and $K_{\rm max} =8$. We consider a noise power of $\sigma^2 = -101$~dBm, a transmitted power at each \ac{bs} of $P=46$~dBm, $\Psi=10$~dB, $\nu=3.5$ and $K_0$ ensures a cell edge \ac{snr} of $20$~dB. Results are obtained by considering that the scheduling process over the entire network generates a total number $T=3\cdot 10^6$ \acp{ipv}.
We consider $D=1$ for both the proposed \ac{mp} and \ac{mq}, i.e., based on the \ac{ipv} estimated at the current time instant, the methods chose an \ac{ipv} at the successive time instant.
We assume as a correct transmission the event in which the predicted \ac{ipv} is greater than the actual one at the successive time instant, whereas otherwise we consider a failure: this models a system where re-transmissions are not allowed. For each sequence of length $T$, we test the reliability of the different solutions by counting the number of events in which the predicted \ac{ipv} is below the actual one, and we consider this value as a proxy of the system reliability. We hence define the reliability of the system as
\begin{equation}
\theta = \frac{\sum_{t=1}^{T-1} \chi(t)}{T-1},
\end{equation}
where $\chi(t)$ is the indicator function of the predicted \ac{ipv} being lower than the actual one, i.e.,
\begin{equation}
\chi(t) =
\begin{cases}
1 \quad \text{if} \, \hat{I}(t+1) < I(t+1); \\
0 \quad \text{otherwise}.
\end{cases}
\end{equation}

For each \ac{ue} we assume that, if a failure happens, the experienced \ac{se} is zero. Therefore the instantaneous \ac{ue} \ac{se} $R(t)$ at time $t$ is computed as
\begin{equation}\label{eq:se}
R(t) = (1-\chi(t))\log_2(1+\hat{\rho}(t)),
\end{equation}
with $\hat{\rho}(t)$ being the predicted \ac{sinr}.

Results are compared against a method based on \cite[Eq. (8)]{pocovi2018}, where the prediction is obtained by low pass filtering the previous estimated \ac{ipv} with the current one, i.e.,
\begin{equation}\label{eq:klaus}
\hat{I}(t+1) = \alpha I(t) + (1-\alpha)\hat{I}(t-1).
\end{equation}
We denote this method as LPP (low-pass prediction). However, note that (\ref{eq:klaus}) has been shown as effective only for \ac{urllc} when \ac{olla} with one re-transmission is available.

Furthermore, we compare our method against the idea proposed in \cite{klein2019}, where authors perform \ac{ipv} prediction based on the marginal \ac{pdf} of the interference. We extend then the proposed approach by applying both the \ac{mp} and \ac{mq} prediction methods to the marginal \ac{pdf} $f_k\left(I(t)\right)$.

\begin{figure}[t]
\includegraphics[width=0.5\textwidth]{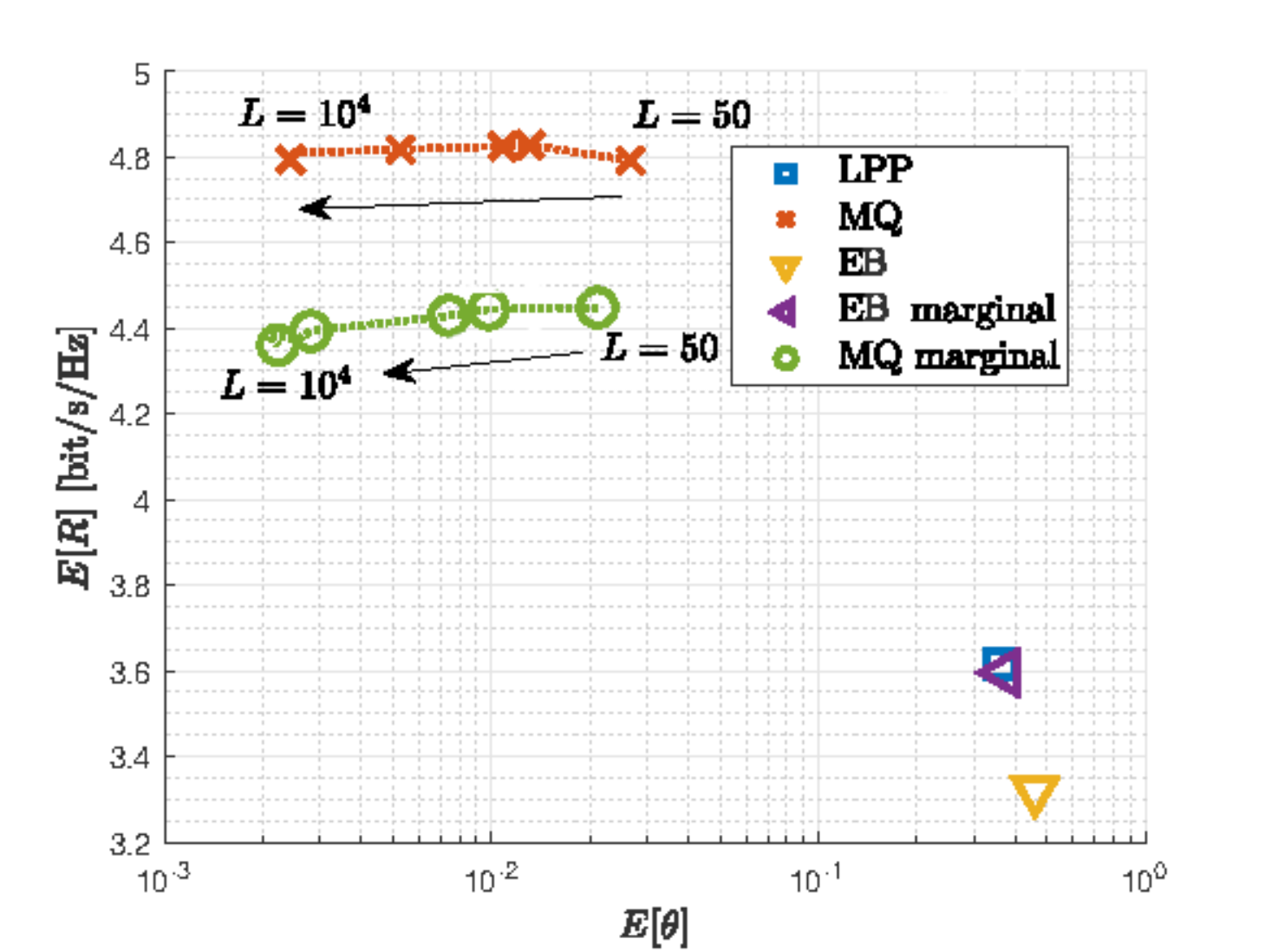} 
\caption{Average \ac{se} $R$ versus average $\theta$ for \ac{rr} with $L=\{50,100,500,5000,10000\}$ and $\Gamma_0 = 10^{-3}$.}
\label{fig:trainRR}
\end{figure}

Fig. \ref{fig:trainRR} shows the average \ac{se} $R$ versus the average $\theta$ for \ac{rr} scheduling with $L=\{50,100,500,5000,10000\}$ and $\Gamma_0 = 10^{-3}$. We first observe that both the \ac{mq} methods on the marginal and conditional \ac{pdf} obtain much better average \ac{se} and $\theta$ when compared to other schemes, as \ac{mq} has been properly designed to control the probability of error without re-transmissions. Furthermore we note that, as $L$ increases, the average $\theta$ decreases for both the \ac{mq} methods on the conditional and on marginal \acp{pdf}. Finally, for each value of $L$ we highlight that the \ac{mq} on conditional \ac{pdf} achieves about $10\%$ higher average rate for the same average $\theta$ when compared to the \ac{mq} on the marginal \ac{pdf}, showing the merits of our proposed approach that exploits time correlation of the interference. Finally, we highlight that both the \ac{mp} methods and LPP do not lead to any better performance when increasing $L$: this is due to the fact that, differently from the \ac{mq} methods, none of these prediction methods has control on the target reliability.

\begin{figure}[t]
\includegraphics[width=0.5\textwidth]{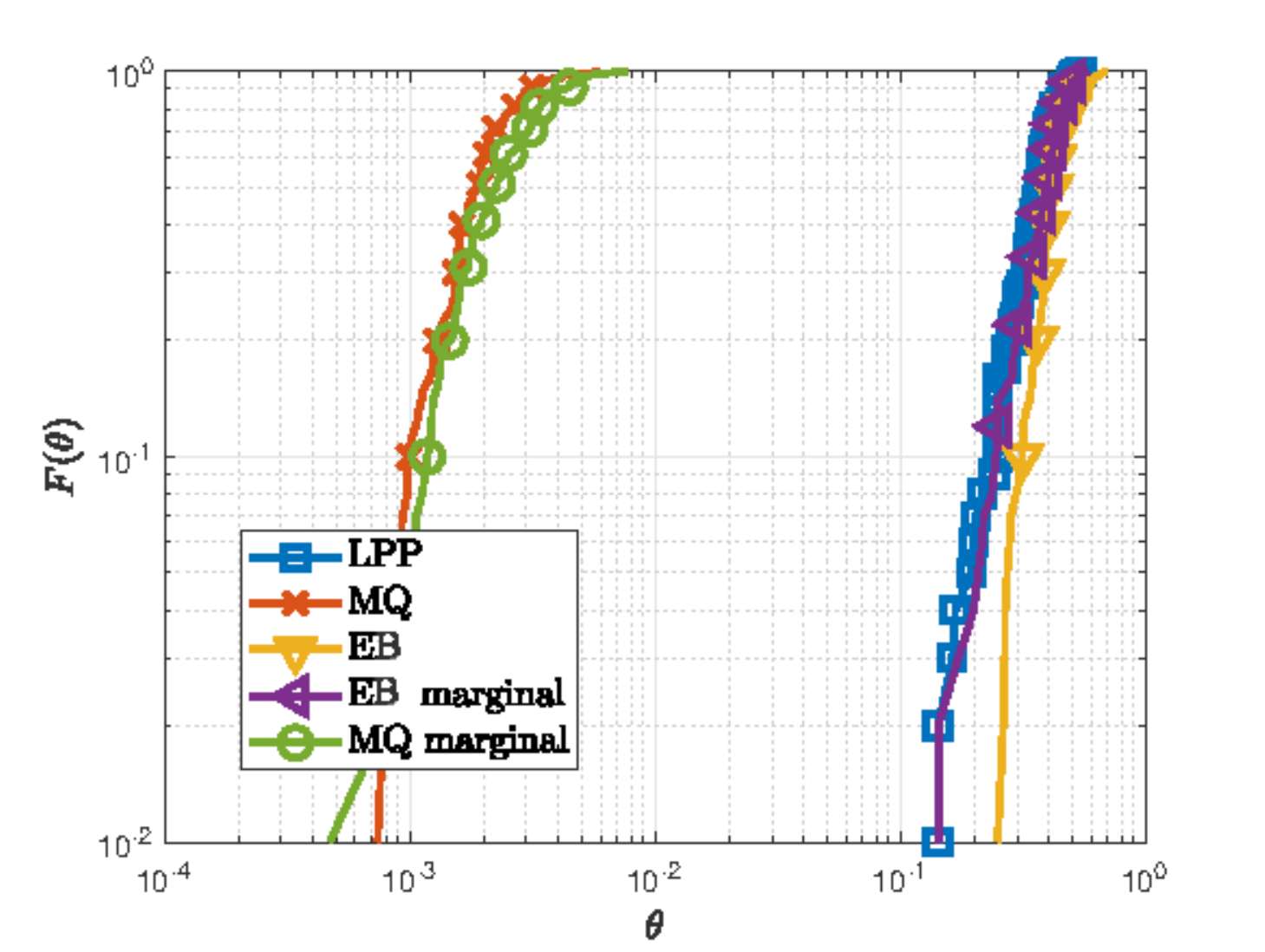}
\caption{\ac{cdf} $F(\theta)$ of $\theta$ for \ac{rr} with $L=5000$ and $\Gamma_0=10^{-3}$.}
\label{fig:cdf}
\end{figure}

Fig. \ref{fig:cdf} shows the \ac{cdf} $F(\theta)$ of $\theta$ for \ac{rr} scheduling with $\Gamma_0=10^{-3}$ and $L=5000$. The \acp{cdf} of the \ac{mq} methods are almost the same, and they present a significant performance gain when compared to the other methods. Furthermore, we observe that the \ac{mq} methods try to meet the target \ac{bler} requirement fixed to $\Gamma_0 = 10^{-3}$ as both \acp{cdf} are centered around that value.

\begin{figure}[t]
\includegraphics[width=0.5\textwidth]{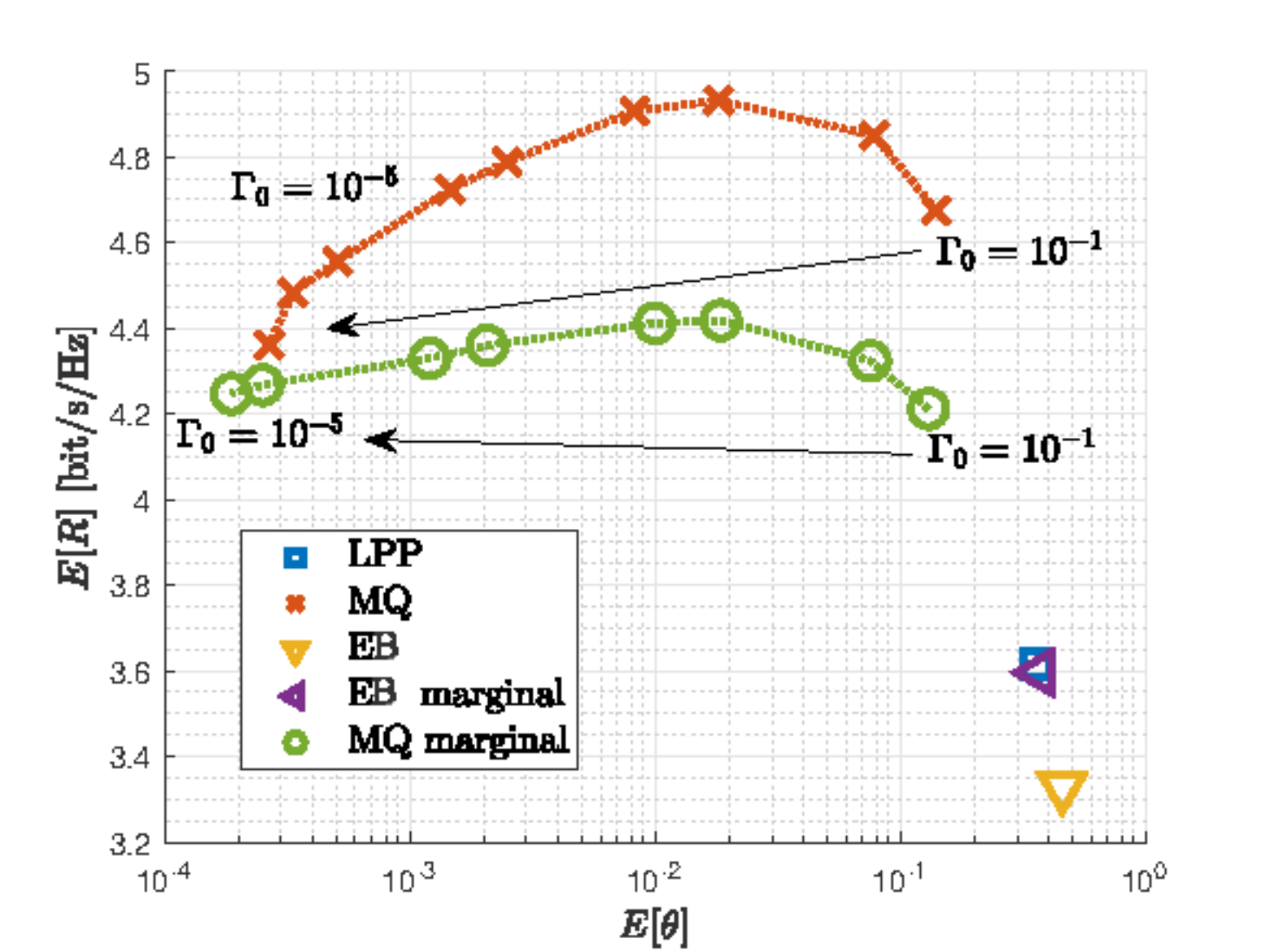}
\caption{Average \ac{se} $R$ versus average $\theta$ for \ac{rr} with $L=5000$ and $\Gamma_0 = \{10^{-1},5\cdot 10^{-2},10^{-2},5\cdot 10^{-3},10^{-3},5\cdot 10^{-4},10^{-4},10^{-5}\}$.}
\label{fig:targetRR}
\end{figure}

Fig. \ref{fig:targetRR} shows the average \ac{se} $R$ versus the average $\theta$ for \ac{rr} scheduling with $L=5000$ and $\Gamma_0 = \{10^{-1},5\cdot 10^{-2},10^{-2},5\cdot 10^{-3},10^{-3},5\cdot 10^{-4},10^{-4},10^{-5}\}$. Similar as before, the proposed \ac{mq} method strongly outperforms LPP and both the \ac{mp} methods in terms of both average \ac{se} and $\theta$. Then, also in this setup with different values of $\Gamma_0$, the \ac{mq} method based on the conditional \ac{pdf} achieves almost the same reliability value $\theta$ of the \ac{mq} based on the marginal \ac{pdf} but providing higher \ac{se}. This further justifies our proposal of using the conditional \ac{pdf}. Finally, as $\Gamma_0$ increases, we notice that the average rate initially increases, while it decreases after a certain point. This is due to the fact that with higher reliability, i.e., lower values of $\Gamma_0$, we tend to be more conservative and waste part of the rate resources that could be used, whereas with lower reliability target, i.e., higher values of $\Gamma_0$, the pre-log factor in (\ref{eq:se}) becomes dominant and therefore the average rate is lowered by the lower success probability. 

\begin{figure}[t]
\includegraphics[width=0.5\textwidth]{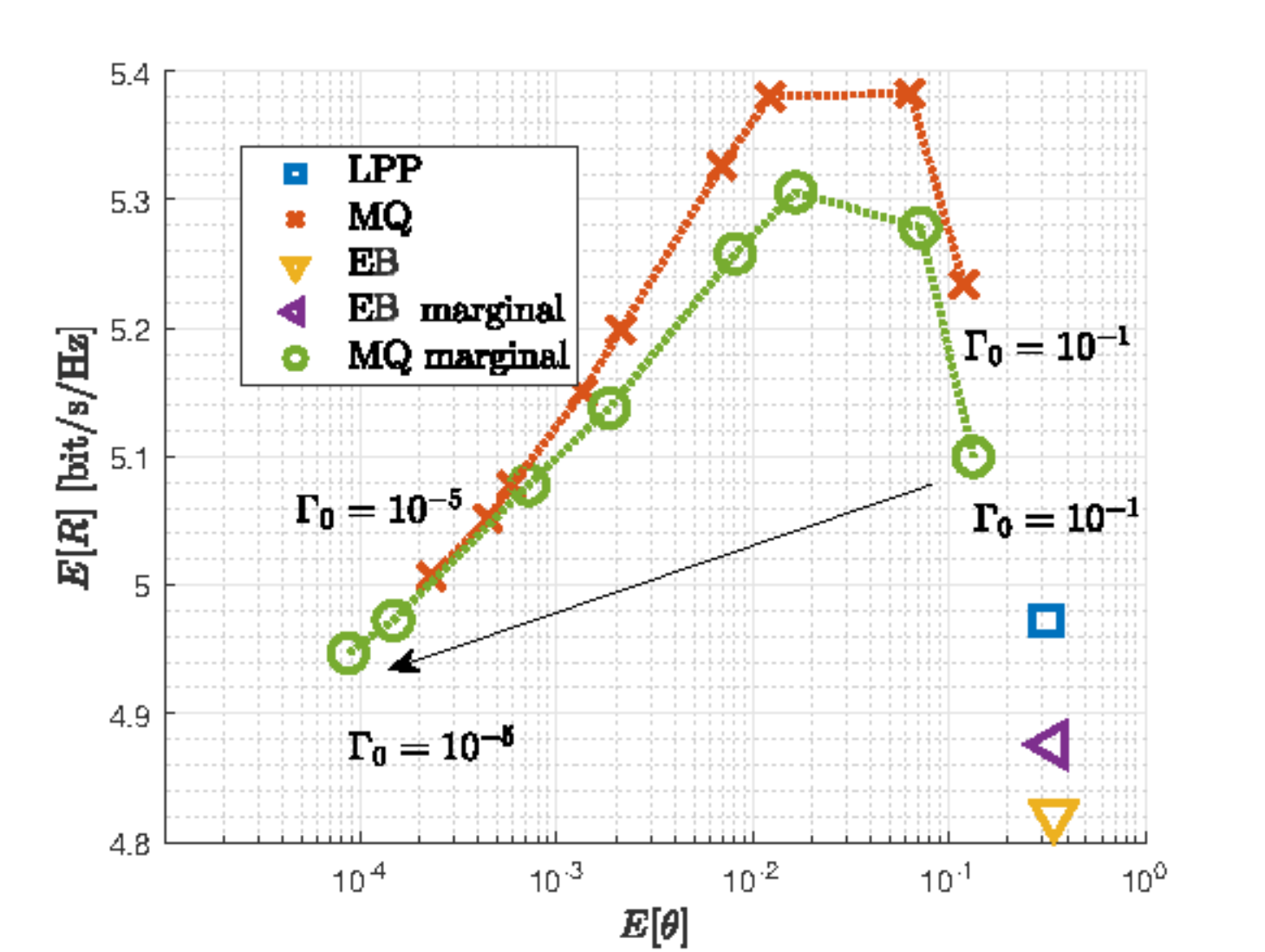}
\caption{Average \ac{se} $R$ versus average $\theta$ for \ac{pfs} with $L=5000$ and $\Gamma_0 = \{10^{-1},5\cdot 10^{-2},10^{-2},5\cdot 10^{-3},10^{-3},5\cdot 10^{-4},10^{-4},10^{-5}\}$.}
\label{fig:targetPFS}
\end{figure}

Although in the considered scenario we have a different number of \acp{ue} per cell, with \ac{rr} there is more time correlation of the interference as the scheduling in each cell is deterministic. Therefore, Fig. \ref{fig:targetPFS} shows the average \ac{se} $R$ versus the average $\theta$ for \ac{pfs} with $L=5000$ and $\Gamma_0 = \{10^{-1},5\cdot 10^{-2},10^{-2},5\cdot 10^{-3},10^{-3},5\cdot 10^{-4},10^{-4},10^{-5}\}$. The increasing and decreasing behavior of the average \ac{se} for increasing values of $\Gamma_0$ is justified by the same considerations done for the \ac{rr} case. Results show that also in the \ac{pfs} case the \ac{mq} solutions perform better than LPP and \ac{mp} methods. Furthermore, we highlight that also in this case \ac{mq} based on the conditional \ac{pdf} is advantageous in terms of average \ac{se} when compared to \ac{mq} based on the marginal \ac{pdf}, although the performance gain is much less than that obtained with \ac{rr}. This shows that, also with \ac{pfs}, there is a certain periodicity in the scheduling, which can still be exploited in order to improve system performance.

\section{Conclusions}
In this paper, we considered the problem of interference prediction for \ac{urllc} traffic. The main motivation behind this work is that, in order to guarantee both the required reliability and at the same time not wasting resources when performing \ac{la}, we must be able to accurately predict the \ac{sinr} in order to choose a proper \ac{mcs}. Considering then that \ac{urllc} with most extreme latency and reliability requirements is characterized by semi-deterministic periodic traffic, we treated interference as a time series. By exploiting the inherent time correlation, we designed two prediction algorithms based on the conditional \ac{pdf}, namely \ac{mp} and \ac{mq}. We compared the proposed solutions against state of the art algorithms, and showed that the \ac{mq} based on the conditional \ac{pdf} is a promising solution for \ac{urllc}, as it allows to control reliability while at the same time optimizing the resources in terms of \ac{se}.

\section{Acknowledgments}
The authors would like to thank their colleagues S. Mandelli, S. Klein, and A. Weber for the numerous helpful discussions.

\balance 

\bibliographystyle{IEEEtran}
\bibliography{bibliography}

\begin{thebibliography}{10}
\providecommand{\url}[1]{#1}
\csname url@samestyle\endcsname
\providecommand{\newblock}{\relax}
\providecommand{\bibinfo}[2]{#2}
\providecommand{\BIBentrySTDinterwordspacing}{\spaceskip=0pt\relax}
\providecommand{\BIBentryALTinterwordstretchfactor}{4}
\providecommand{\BIBentryALTinterwordspacing}{\spaceskip=\fontdimen2\font plus
\BIBentryALTinterwordstretchfactor\fontdimen3\font minus
  \fontdimen4\font\relax}
\providecommand{\BIBforeignlanguage}[2]{{%
\expandafter\ifx\csname l@#1\endcsname\relax
\typeout{** WARNING: IEEEtran.bst: No hyphenation pattern has been}%
\typeout{** loaded for the language `#1'. Using the pattern for}%
\typeout{** the default language instead.}%
\else
\language=\csname l@#1\endcsname
\fi
#2}}
\providecommand{\BIBdecl}{\relax}
\BIBdecl

\bibitem{3gpp.22.104}
3GPP, ``{Service requirements for cyber-physical control applications in
  vertical domains},'' {3rd Generation Partnership Project (3GPP)}, TS
  {22.104}, 2019.

\bibitem{durisi2016}
G.~Durisi, T.~Koch, and P.~Popovski, ``Toward massive, ultrareliable, and
  low-latency wireless communication with short packets,'' \emph{Proc. of the
  IEEE}, vol. 104, no.~9, pp. 1711--1726, 2016.

\bibitem{li20175g}
C.-P. Li, J.~Jiang, W.~Chen, T.~Ji, and J.~Smee, ``{5G} ultra-reliable and
  low-latency systems design,'' in \emph{2017 European Conf. on Networks and
  Commun. (EuCNC)}.\hskip 1em plus 0.5em minus 0.4em\relax IEEE, 2017, pp.
  1--5.

\bibitem{bennis2018}
M.~Bennis, M.~Debbah, and H.~V. Poor, ``Ultrareliable and low-latency wireless
  communication: Tail, risk, and scale,'' \emph{Proc. of the {IEEE}}, vol. 106,
  no.~10, pp. 1834--1853, 2018.

\bibitem{shariatmadari2016}
H.~Shariatmadari, Z.~Li, M.~A. Uusitalo, S.~Iraji, and R.~J{\"a}ntti, ``Link
  adaptation design for ultra-reliable communications,'' in \emph{2016 {IEEE}
  International Conf. on Commun. (ICC)}.\hskip 1em plus 0.5em minus 0.4em\relax
  IEEE, 2016, pp. 1--5.

\bibitem{sampath1997}
A.~Sampath, P.~S. Kumar, and J.~M. Holtzman, ``On setting reverse link target
  {SIR} in a {CDMA} system,'' in \emph{1997 {IEEE} 47th Vehicular Tech. Conf.
  Technology in Motion}, vol.~2.\hskip 1em plus 0.5em minus 0.4em\relax IEEE,
  1997, pp. 929--933.

\bibitem{pocovi2018}
G.~Pocovi, K.~I. Pedersen, and P.~Mogensen, ``Joint link adaptation and
  scheduling for {5G} ultra-reliable low-latency communications,'' \emph{{IEEE}
  Access}, vol.~6, pp. 28\,912--28\,922, 2018.

\bibitem{oruthota2016}
U.~Oruthota, F.~Ahmed, and O.~Tirkkonen, ``Ultra-reliable link adaptation for
  downlink {MISO} transmission in {5G} cellular networks,'' \emph{Information},
  vol.~7, no.~1, p.~14, 2016.

\bibitem{levanen2015}
T.~Levanen, J.~Venalainen, and M.~Valkama, ``Interference analysis and
  performance evaluation of {5G} flexible-{TDD} based dense small-cell
  system,'' in \emph{2015 {IEEE} 82nd Vehicular Tech. Conf.
  (VTC2015-Fall)}.\hskip 1em plus 0.5em minus 0.4em\relax IEEE, 2015, pp. 1--7.

\bibitem{taranetz2015}
M.~Taranetz and M.~Rupp, ``A circular interference model for heterogeneous
  cellular networks,'' \emph{{IEEE} Trans. on Wireless Commun.}, vol.~15,
  no.~2, pp. 1432--1444, 2015.

\bibitem{feng2019}
K.~Feng and M.~Haenggi, ``On the location-dependent {SIR} gain in cellular
  networks,'' \emph{{IEEE} Wireless Commun. Letters}, vol.~8, no.~3, pp.
  777--780, 2019.

\bibitem{zhuang2011}
Y.~Zhuang, Y.~Luo, L.~Cai, and J.~Pan, ``A geometric probability model for
  capacity analysis and interference estimation in wireless mobile cellular
  systems,'' in \emph{2011 {IEEE} Global Telecom. Conf. (GLOBECOM)}.\hskip 1em
  plus 0.5em minus 0.4em\relax IEEE, 2011, pp. 1--6.

\bibitem{balanis2016}
C.~A. Balanis, \emph{Antenna theory: analysis and design}.\hskip 1em plus 0.5em
  minus 0.4em\relax John wiley \& sons, 2016.

\bibitem{tse2005}
D.~Tse and P.~Viswanath, \emph{Fundamentals of wireless communication}.\hskip
  1em plus 0.5em minus 0.4em\relax Cambridge university press, 2005.

\bibitem{viswanath2002}
P.~Viswanath, D.~N.~C. Tse, and R.~Laroia, ``Opportunistic beamforming using
  dumb antennas,'' \emph{{IEEE} Trans. on Inform. Theory}, vol.~48, no.~6, pp.
  1277--1294, 2002.

\bibitem{sisinni2016}
E.~Sisinni and F.~Tramarin, ``Isochronous wireless communication system for
  industrial automation,'' in \emph{Industrial Wireless Sensor Networks}.\hskip
  1em plus 0.5em minus 0.4em\relax Elsevier, 2016, pp. 167--188.

\bibitem{holma2012}
H.~Holma and A.~Toskala, \emph{LTE advanced: 3GPP solution for
  IMT-Advanced}.\hskip 1em plus 0.5em minus 0.4em\relax John Wiley \& Sons,
  2012.

\bibitem{botev2010}
Z.~I. Botev, J.~F. Grotowski, D.~P. Kroese \emph{et~al.}, ``Kernel density
  estimation via diffusion,'' \emph{The annals of Statistics}, vol.~38, no.~5,
  pp. 2916--2957, 2010.

\bibitem{klein2019}
S.~Klein and S.~Mandelli, ``Link adaptation in telecommunication systems,''
  2019, filed patent application.

\end{thebibliography}

\end{document}